\documentstyle[12pt,equation,epsfig]{article}
\begin{document}
\newcommand{\be}{\begin{equation}}
\newcommand{\ben}{\begin{subequations}}
\newcommand{\een}{\end{subequations}}
\newcommand{\beq}{\begin{eqalignno}}
\newcommand{\eeq}{\end{eqalignno}}
\newcommand{\ee}{\end{equation}}
\newcommand{\epem}{\mbox{$e^+ e^-$}}
\newcommand{\tanb}{\mbox{$\tan \! \beta$}}
\newcommand{\mhpl}{\mbox{$m_{H^+}$}}
\newcommand{\stau}{\mbox{$\widetilde \tau$}}
\newcommand{\tchi}{\mbox{$\tilde \chi$}}
\newcommand{\ttau}{\tilde \tau}
\renewcommand{\thefootnote}{\fnsymbol{footnote}}

\pagestyle{empty}
\begin{flushright}
APCTP 98--14 \\
YUMS 98--9\\
May 1998\\
\end{flushright}
\vspace*{2cm}
\begin{center}
{\Large \bf CP--Violation through Scalar Tau Oscillation}\\
\vspace*{6mm}
Seong--Youl Choi$^1$ and Manuel Drees$^2$ \\
$^1${\it Dept. of Physics, Yonsei Univ., Seoul 120--749, Korea} \\
$^2${\it APCTP, 207--43 Cheongryangryi--dong, Tongdaemun--gu, Seoul
130--012, Korea}
\end{center}

\vspace*{1cm}
\begin{abstract}
We point out that oscillation between the two scalar tau (\stau) mass
eigenstates can give rise to CP--violation if some parameters appearing
in the stau or chargino/neutralino mass matrices are complex. If 
$\stau^+$ and $\stau^-$ decay into different charginos or neutralinos,
rate asymmetries as large as 20\% are possible. If both staus decay
directly into $\tau+$LSP, CP--violation can in principle still be observed
through an energy asymmetry of the $\tau$ decay products, but this asymmetry
never exceeds the percent level. Even the rate asymmetries become small if
the mass splitting between the two stau mass eigenstates is larger than 1\%.
\end{abstract}

\clearpage
\pagestyle{plain}
\setcounter{page}{1}

\section*{1) Introduction}

The currently most widely studied extensions of the Standard Model
(SM) involve Supersymmetry (SUSY). Superparticles can stabilize the
gauge hierarchy \cite{1}, and allow for the Grand Unification of all
three gauge groups of the SM \cite{2}. Fortunately both of these
statements remain true in (softly) broken SUSY models, which satisfy
experimental constraints from the unsuccessful searches for
superparticles, most notably at LEP and the Tevatron
\cite{3}. Unfortunately neither of these arguments tells us anything
about the dynamics, or even the scale, of supersymmetry breaking as
long as (most) sparticles are not (much) heavier than 1 TeV.  In
phenomenological analyses it is therefore preferable to take as
general an approach to SUSY breaking as possible, i.e. to simply
parameterize it in terms of soft breaking operators. Many of these
parameters can be complex, and therefore can give rise to many new
CP--violating effects.

In this note we study CP--violation in the scalar tau (\stau) sector
at a future (linear) \epem\ collider. There are several reasons to
assume that CP--violation might be much more prominent for \stau\ than
for first or second generation sleptons. In the absence of generation
mixing, all nontrivial phases in the
lepton--slepton--chargino--neutralino part of Lagrangian of the
Minimal Supersymmetric Standard Model (MSSM) can be shifted into terms
involving lepton masses and Yukawa couplings. Moreover, present
experimental bounds \cite{3} on the CP--odd electric and weak dipole
moments of the $\tau$ lepton are too weak to significantly constrain
CP--violation in the \stau\ sector, in sharp contrast to the case of
(s)electrons and, to a lesser extent, (s)muons. In addition, the short
lifetime of the $\tau$ lepton allows one to determine its
polarization, at least in a statistical sense, by measuring the energy
distribution of its visible decay products; this opens the possibility
to construct new CP--odd observables. Finally, unlike \stau's, because
of their small Yukawa couplings first and second generation sleptons
could have masses in the tens of TeV range without destabilizing the
gauge hierarchy \cite{4}; indeed, this is one possible \cite{5}
(although not very attractive \cite{6}) solution of the ``SUSY flavor
problem''.

In spite of these merits, CP--violation in \stau\ pair events can only
be observable if the two \stau\ mass eigenstates are closely
degenerate. The reason is that the two produced \stau\ sleptons, being
scalar particles, decay independently of each other. All CP--odd
variables are therefore proportional to the product of a CP--odd phase
and a second, CP--even phase; this is analogous to the more familiar
$K^0 - \overline{K^0}$ and $B^0 - \overline{B^0}$ systems. In the case
at hand this second phase can only come from the Breit--Wigner
propagators of the \stau\ sleptons,\footnote{In general there can also
be other dispersive phases, e.g. due to loop corrections to the \stau\
decay vertices. However, they are too small to be of much use.} or,
more specifically, from the interference of two {\em different} \stau\
propagators. These interference contributions, which can be understood
in terms of oscillation between the two \stau\ mass eigenstates, will
become very small if the mass difference is much larger than the
(average) decay width of the two \stau\ states.

In this respect the situation is somewhat similar to the case of
CP--violation through slepton flavor oscillation \cite{7,7a} (see also
refs.\cite{7b} for earlier, related work). There are two differences,
however. First, the two \stau\ current eigenstates do not carry
different quantum numbers that are conserved in the SM, while sleptons
of different generations obviously do; even detecting the presence of
mixing is therefore not entirely straightforward in our case
\cite{8}. Second, the two \stau\ current eigenstates have different
$SU(2) \times U(1)_Y$ quantum numbers. According to our current
understanding of SUSY breaking there is therefore no good reason to
assume that the soft breaking masses for these states should be close
to each other. On the other hand, nothing forbids such an
``accidental'' near--degeneracy, either. It might therefore be useful
to point out that such a degeneracy can lead to interesting new
phenomena.

The rest of this article is organized as follows. In the next Section
we discuss rate asymmetries, which can occur if the positive and
negative \stau\ decay into final states that are not CP--conjugates of
each other.  We find that under favorable circumstances these
asymmetries could be detectable at future \epem\ colliders. In Sec.~3
we discuss the pion energy asymmetry that can result from $\stau
\rightarrow \tau \rightarrow \pi$ decays if both \stau\ decay into the
same neutralino. We find this asymmetry to be too small to be
detectable with the currently foreseen luminosity of next--generation
colliders, at least within the MSSM.  Finally, Sec.~4 is devoted to a
summary and conclusions.

\section*{2) Rate Asymmetries}
We begin our discussion with a treatment of rate asymmetries in processes
of the kind 
\be \label{e0}
\epem \rightarrow \stau^- \stau^+ \rightarrow \left( \tchi f
\right)_a^- \left ( \tchi f \right)_b^+.
\ee
Here $a, b$ labels the combination of a $\tau$ with a neutralino
$\tchi_n^0$ ($n=1$ to 4 in the MSSM) or a $\nu_\tau$ with a chargino
$\tchi_m^\pm$ ($m=1$ or 2).  It is crucial to include all combinations
$\stau_i^- \stau_j^+$ in the intermediate state, where $i,j \in
\{1,2\}$ labels the \stau\ mass eigenstates. They are determined by
the mass matrix in $\stau_L, \stau_R$ basis:
\be \label{e1}
{\cal M}^2_{\ttau} = \mbox{$ \left( \begin{array}{cc}
m^2_{\ttau_L} & - m_\tau \left( A_\tau + \mu \tanb \right) \\
- m_\tau \left( A^*_\tau + \mu^* \tanb \right) & m^2_{\ttau_R}
\end{array} \right). $}
\ee
We have absorbed the $D-$ term contributions to the diagonal entries,
as well as the supersymmetric contributions $+ m_\tau^2$, into the
definition of $m^2_{\ttau_L}$ and $m^2_{\ttau_R}$. Note that we allow
the soft breaking parameter $A_\tau$ as well as the supersymmetric
higgsino mass parameter $\mu$ to be complex. Defining the \stau\ mass
eigenstates through $\left( \stau_1, \stau_2 \right) = \left( \stau_L,
\stau_R \right) U^T_{\ttau}$, with
\be \label{e2}
U_{\ttau} = \mbox{$ \left( \begin{array}{cc} 
\cos \! \theta_{\ttau} & e^{i \phi_{\ttau}} \sin \! \theta_{\ttau} \\
-e^{-i \phi_{\ttau}} \sin \! \theta_{\ttau} & \cos \! \theta_{\ttau}
\end{array} \right), $}
\ee
we have
\ben \label{e3} \beq
\phi_{\ttau} &= - \arg \left( A_\tau + \mu \tanb \right);
\label{e3a} \\
\tan \! \theta_{\ttau} &= \frac {m^2_{\ttau_1} - m^2_{\ttau_L} }
{ m_\tau \left| A_\tau + \mu \tanb \right| } .
\label{e3b}
\eeq \een
The eigenvalues $m^2_{{\ttau}_{1,2}}$ are not sensitive to the phase
$\phi_{\ttau}$.

Since there are two \stau\ mass eigenstates, the squared matrix element,
or the cross section, for process (\ref{e0}) contains a total of 16
terms:
\beq \label{e4}
\sigma  \left( \epem \rightarrow \left[ \tchi f \right]_a^- \left[ \tchi f
\right]_b^+ \right) &= \frac {e^4} { (2 \pi)^3 s} \cdot 
\frac {1} {3 \cdot 2^{10}} 
 \cdot \sum_{i,j,k,l} \lambda^{3/2} \left( 1, 
\frac{ m^2_{\ttau_{ik}} } {s}, \frac{ m^2_{\ttau_{jl}} } {s} \right)
\left( P_L^{\rm eff} c^L_{ij} c^{L*}_{kl} + P_R^{\rm eff} c^R_{ij}
c^{R*}_{kl} \right) \nonumber \\
& \cdot A_{ik} A_{jl} \lambda^{1/2} \left( 1,
\frac {m^2_{\tilde \chi_a}} { m^2_{\ttau_{ik}} },
\frac {m^2_{f_a}} { m^2_{\ttau_{ik}} } \right)
\lambda^{1/2} \left( 1, \frac {m^2_{\tilde \chi_b}} { m^2_{\ttau_{jl}} },
\frac {m^2_{f_b}} { m^2_{\ttau_{jl}} } \right) D^a_{ik} D^{b*}_{jl},
\eeq
where $e$ is the QED gauge coupling, $s$ is the squared \epem\ c.m.
energy, and $\lambda(a,b,c) = (a+b-c)^2 - 4 ab$. The first two factors
under the sum describe $\epem \rightarrow \stau^- \stau^+$
production. We have allowed for non--degenerate \stau\ states by
introducing the average masses $m_{\ttau_{ik}} = \left( m_{\ttau_i} +
m_{\ttau_k} \right)/2$; nevertheless eq.(\ref{e4}) will not describe
the (small) interference terms very well if $m_{\ttau_2} -
m_{\ttau_1}$ is large. The coefficients $P_L^{\rm eff} = \left( 1 -
P_{e^-} \right) \left( 1 + P_{e^+} \right)$ and $P_R^{\rm eff} =
\left( 1 + P_{e^-} \right) \left( 1 - P_{e^+} \right)$ describe the
longitudinal polarization of the $e^\pm$ beams. The effective
couplings $c^L_{ij}, \ c^R_{ij}$ are given by:
\ben \label{e5} \beq
c^L_{ij} &= \delta_{ij} + \frac { 1 - 2 \sin^2 \theta_W} {2 \sin^2
\theta_W \cos^2 \theta_W} \cdot \frac {s} {s-M_Z^2} c^Z_{ij};
\label{e5a} \\
c^R_{ij} &= \delta_{ij} - \frac {1} {\cos^2 \theta_W} \cdot
\frac {s} {s-M_Z^2} c^Z_{ij},
\label{e5b}
\eeq \een
where the $Z \stau^-_i \stau^+_j$ couplings $c^Z_{ij}$ are defined as
\be \label{e6}
c^Z_{11} = \frac {1}{2} \cos^2 \theta_{\ttau} - \sin^2 \theta_W; \ \ 
c^Z_{22} = \frac {1}{2} \sin^2 \theta_{\ttau} - \sin^2 \theta_W; \ \
c^Z_{12} = c^{Z*}_{21} = - \frac{1}{4} \sin \left( 2 \theta_{\ttau} \right)
e^{i \phi_{\ttau}}.
\ee

The $A_{ik}$ factors in eq.(\ref{e4}) describe the convolution of two
Breit--Wigner propagators. We follow ref.\cite{7} in using the narrow
width approximation, but allow for different decay widths of the two 
\stau\ mass eigenstates:
\ben \label{e8} \beq
\Re e \left( A_{ik} \right) &= \frac {1} { m_{\ttau_{ik}} 
\Gamma_{\ttau_{ik}} } \cdot \frac {1} {1 + \frac { \left( m_{\ttau_i} -
m_{\ttau_k} \right)^2 } { \Gamma_{\ttau_i} \Gamma_{\ttau_k} } }
\left( 1 + \frac {2 x_{ik} } {\pi} \log \frac {m_{\ttau_k}} {m_{\ttau_i}}
\right);
\label{e8a} \\
\Im m \left( A_{ik} \right) &= -\frac {1} { m_{\ttau_{ik}} 
\Gamma_{\ttau_{ik}} } \cdot \frac {x_{ik}} {1 + x^2_{ik}},
\label{e8b}
\eeq \een
where we have introduced the average widths $\Gamma_{\ttau_{ik}} =
\left( \Gamma_{\ttau_i} + \Gamma_{\ttau_k} \right)/2$ as well as the
``\stau\ oscillation parameter''
\be \label{e9}
x_{ik} = \frac { m_{\ttau_i} - m_{\ttau_k} } { \Gamma_{\ttau_{ik}} }.
\ee
The second (logarithmic) term in eq.(\ref{e8a}) has been introduced to
improve agreement with a numerical convolution of two propagators for
$m_{\ttau_2} \geq 1.05 m_{\ttau_1}$; however, we will see that in this 
region of parameter space CP--violating effects are already very small.
Notice that $x_{ki} = -x_{ik}$, i.e. $A_{ki} = A^*_{ik}$.

The last factors in eq.(\ref{e4}) describe $\stau \rightarrow \left( \tchi
f \right)_a$ decays. In particular, $D^a_{ik}$ is given by:
\be \label{e10}
D^a_{ik} = \left( m^2_{\ttau_{ik}} - m^2_{\tilde \chi_a} - m^2_{f_a}
\right) \left( L_{ia} L^*_{ka} + R_{ia} R^*_{ka} \right)
%\nonumber \\
- 2 m_{f_a} m_{\tilde \chi_a} \left( L_{ia} R^*_{ka} + R_{ia} L^*_{ka}
\right).
\ee
Here, $L_{ia}$ and $R_{ia}$ are the $\stau_i (\tchi f)_a$ couplings
for left-- and right--handed $f_a$, respectively. They can be computed
easily from the $\tilde{f} f \tchi$ interactions listed in ref.\cite{9},
together with eqs.(\ref{e2}) and (\ref{e3}) describing \stau\ mixing;
of course, care must be taken to allow for complex chargino and neutralino
mixing matrices. Note that $m_{f_a} = R_{ia} = 0$ for $\stau^- \rightarrow
\tchi^- \nu_\tau$ decays.

CP--odd rate asymmetries can be defined as
\be \label{e11}
A( \tchi_a \tchi_b) = \frac { \sigma (a^- b^+) -
\sigma (b^- a^+) } { \sigma (a^- b^+) + \sigma (b^- a^+) },
\ee
where we have used the short--hand notation $a^\pm = (\tchi f)^\pm_a$.
From eq.(\ref{e4}) one finds
\beq \label{e12}
A( \tchi_a \tchi_b ) &\propto \sum_{i,j,k,l} A_{ik} A_{jl}
\left( c_{ij} c^*_{kl} D^a_{ik} D^{b*}_{jl} - 
       c_{ij} c^*_{kl} D^b_{ik} D^{a*}_{jl} \right)
\nonumber \\
&= \sum_{i,j,k,l} A_{ik} A_{jl} \left( c_{ij} c^*_{kl} D^a_{ik} D^{b*}_{jl}
- c_{ji} c^*_{lk} D^{b}_{jl} D^{a*}_{ik} \right)
\nonumber \\
&= 2i \sum_{i,j,k,l} A_{ik} A_{jl} \Im m \left( c_{ij} c^*_{kl} D^a_{ik}
D^{b*}_{jl} \right)
\nonumber \\
&= -2 \sum_{i,j,k,l} \Im m \left( A_{ik} A_{jl} \right)
\Im m \left( c_{ij} c^*_{kl} D^a_{ik} D^{b*}_{jl} \right),
\eeq
where we have used the abbreviation $c_{ij} c^*_{kl} = P_L^{\rm eff}
c^L_{ij} c^{L*}_{kl} + P_R^{\rm eff} c^R_{ij} c^{R*}_{kl}$. In the
second term of the second line of eq.(\ref{e12}) we have simply
relabelled the indices as $i \leftrightarrow j, \ k \leftrightarrow
l$, and in the third line we have used $c_{ji} = c^*_{ij}$. In the
last step we have symmetrized under $i \leftrightarrow k, \ l
\leftrightarrow j$ to show that the final result is real. (Recall that
$A_{ki} = A^*_{ik}$.)

Since there are four neutralinos and two charginos in the MSSM, one
can in principle define 15 independent rate asymmetries. We found that
for \stau\ masses within reach of a next--generation \epem\ collider
operating at $\sqrt{s} \simeq 500$ GeV the two most readily accessible
asymmetries, $A( \tchi^0_1 \tchi_1^+ )$ and $A ( \tchi^0_1 \tchi^0_2
)$, are also the most promising ones. We see from eqs.(\ref{e12}) and
(\ref{e8b}) that the rate asymmetries can only be sizable if the
oscillation parameter $x_{12}$ is not much larger (nor much smaller)
than 1. Since the total width of \stau\ eigenstates with mass $\leq
200$ GeV does not exceed 1 GeV, it is clear that a sizable effect can
only be expected if the mass splitting is quite small. A numerical
scan of the parameter space reveals that the rate asymmetries become
maximal for $m_{\ttau_L} = m_{\ttau_R}, \ \left|A_\tau \right| = |\mu
\tanb|$, and $\arg (A_\tau) \simeq \arg(\mu) - 3.2$; the last two
conditions ensure that the two contributions to the off--diagonal
entries of the \stau\ mass matrix cancel approximately (but not
exactly). The condition for $\left| A_\tau \right|$ can usually only
be satisfied for $\tanb \leq 5$, at least if we insist that the scalar
potential should not have a deeper lying minimum where the \stau\
fields get non-vanishing vacuum expectation values.

In Fig.~1 we show a scatter plot in the plane spanned by the two
``effective'' asymmetries $\hat{A}( \tchi_1^0 \tchi_1^+ )$ and
$\hat{A}( \tchi_1^0 \tchi_2^0 )$, where we have fixed $m_{\ttau_L} =
m_{\ttau_R} = 200$ GeV, picked random combinations of $M_2, \ \mu$ and
\tanb, and fixed the absolute value and phase of $A_\tau$ as described
above. We have assumed that the $SU(2)$ and $U(1)_Y$ gaugino masses
unify, which allows us to set both of their phases to zero without
loss of generality;\footnote{In general, $\arg(M_1) - \arg(M_2), \
\arg(\mu) + \arg(M_2)$ and $\arg(A_\tau) + \arg(M_2)$ are the
rephasing invariant CP--violating phases relevant to our problem. In
our case the phase of $A_\tau$ only enters through the \stau\ mixing
phase $\phi_{\ttau}$ defined in eq.(\ref{e3a}); however, the phase of
$\mu$ also appears independently in the chargino and neutralino mass
matrices.} we have checked that introducing a relative phase between
these two soft breaking parameters does not change the result
qualitatively. The ``effective'' asymmetries are defined as
\be \label{e13}
\hat{A} ( \tchi_a \tchi_b ) = A ( \tchi_a \tchi_b )
\cdot \sqrt{ \sigma\left( a^- b^+ \right) + \sigma \left( a^+ b^- \right) }.
\ee
They directly determine the luminosity needed to experimentally detect
an asymmetry with a significance of $N_\sigma$ standard deviations:
\be \label{e14}
\epsilon {\cal L}(N_\sigma) = \left( \frac {N_\sigma} { \hat{A}
(\tchi_a \tchi_b) } \right)^2,
\ee
where $\epsilon$ is the overall efficiency. Fig.~1 shows that the
anticipated luminosity of 500 GeV colliders, ${\cal L} \simeq 20$ to
50 fb$^{-1}/$yr, should be sufficient to probe at least some regions of
parameter space, if the more complicated \stau\ decay modes can be
reconstructed with an efficiency not far below the value of $\sim 50\%$
found in refs.\cite{8} for $\tau^+ \tau^- \tchi_1^0 \tchi_1^0$ final
states where both $\tau$ leptons decay hadronically.

We observe some correlation between the two effective asymmetries if
they have the same sign.\footnote{Since all asymmetries change sign
when the sign of all CP--odd phases in the Lagrangian is flipped, we
only show results for $\hat{A}( \tchi_1^0 \tchi_1^+ ) > 0$.}  The
reason is that an enhancement of the $\tchi_1^0$ mode in the
negatively charged ($\stau^-$) channel, or a suppression of this mode
in the positively charged ($\stau^+$) channel, affects both
asymmetries in the same way. On the other hand, the much weaker
correlation between $\hat{A}( \tchi_1^0 \tchi_1^+ )$ and $\hat{A} (
\tchi_1^0 \tchi_2^0 )$ for $A ( \tchi_1^0 \tchi_2^0 ) < 0$ can be
explained by the observation that both asymmetries would vanish in the
limit of vanishing $\tau$ Yukawa coupling, or vanishing gauge
couplings. The final state chargino/neutralino therefore needs
significant gaugino--higgsino mixing if the asymmetries are to be
sizable. Since the parameters that describe the chargino mass matrix
also appear in the neutralino mass matrix, large gaugino--higgsino
mixing in one sector tends to lead to large mixing in the other sector
as well. Finally, we note that the scenarios with the largest
effective asymmetries tend to have a fairly small phase of the $\mu$
parameter, of order 0.1 or less; this makes it a little easier to
satisfy constraints from the electric dipole moments of the neutron
and electron, either by choosing large values for first generation
sfermion masses, or by tuning the phases of the relevant $A$
parameters.

In Fig.~2 we show the reduction of the asymmetry $A ( \tchi_1^0
\tchi_1^+ )$ as we deviate from near--degeneracy of the two \stau\
mass eigenstates. As discussed above, three conditions need to be
satisfied for this near--degeneracy to occur, which means that one can
deviate from it in three different directions: One can introduce some
splitting between the diagonal entries of the mass matrix (\ref{e1}),
or one can vary the phase or absolute value of $A_\tau$. These three
directions are explored in the solid, short dashed and long dashed
curves of Fig.~2, respectively, starting from a choice of parameters
leading to a large asymmetry ($\sim 20\%$). Along the solid line,
$\left| m_{\ttau_R} - m_{\ttau_L} \right| \simeq m_{\ttau_2} -
m_{\ttau_1}$, while $|A_\tau|$ varies by more than 100 GeV along the
long dashed curve; the impression that $A( \tchi_1^0 \tchi_1^+ )$
depends more sensitively on $|A_\tau|$ than on $m_{\ttau_L} -
m_{\ttau_R}$ is therefore somewhat misleading. Finally, the phase of
$A_\tau$ varies by about $\pi/10$ along the short dashed curve. Fig.~2
clearly illustrates that rate asymmetries will be too small to be
detectable at a next--generation \epem\ collider if the two
eigenvalues of the \stau\ mass matrix (\ref{e1}) differ by more than
1\%.

\section*{3) Energy Asymmetries}

If $\stau^+$ and $\stau^-$ decay into charge--conjugate final states
no rate asymmetry can be measured. This will be true in particular if
$\tau \tchi_1^0$ is the only accessible \stau\ decay mode. In this
case one can still construct a CP--odd observable involving the spins
of the two $\tau$ leptons. While these spins are not directly
measurable, they affect the energy distributions of the visible $\tau$
decay products.  Here we study this effect using the simplest $\tau$
decay mode, $\tau^\pm \rightarrow \pi^\pm \nu_\tau$, which also has
the highest sensitivity to the $\tau$ polarization.

Since the CP--odd observable we want to study here depends on
kinematical quantities, we need an expression for the differential
cross section, rather than the total one; this also allows us to
implement acceptance cuts that are needed to isolate \stau\ pair
events from SM backgrounds.  It can be written as:
\beq \label{e15}
d \sigma( \epem &\rightarrow \stau^- \stau^+ \rightarrow \tau^- \tau^+
\tchi_1^0 \tchi_1^0 \rightarrow \pi^- \pi^+ \nu_\tau \bar \nu_\tau
\tchi_1^0 \tchi_1^0 ) = 
\nonumber \\ 
&\frac {1} {(2 \pi)^5} \frac {e^4} {2^{12} s} \cdot \left [ B(\tau^-
\rightarrow \pi^- \nu_\tau ) \right]^2 \sin^2 \Theta_{\ttau}
d \cos \Theta_{\ttau} \sum_{i,j,k,l} \lambda^{3/2} \left( 1,
\frac { m^2_{\ttau_{ik}} } {s}, \frac { m^2_{\ttau_{jl}} } {s} \right)
\nonumber \\ 
& \cdot \frac {1} { m_{\ttau_{ik}} m_{\ttau_{jl}} } d \Omega^*_{\pi^-}
d E^*_{\pi^-}  d \Omega^*_{\pi^+}d E^*_{\pi^+} \left( P_L^{\rm eff}
c^L_{ij} c^{L*}_{kl} + P_R^{\rm eff} c^R_{ij} c^{R*}_{kl} \right)
%\nonumber \\ 
\cdot A_{ik} A_{jl} D^\pi_{ak} D^{\pi *}_{jl}.
\eeq
Here, $\Theta_{\ttau}$ is the angle of $\stau^-$ with respect to the
$e^-$ beam direction in the lab frame, while the starred variables
$\Omega^*_{\pi^\pm}$ and $E^*_{\pi^\pm}$ refer to the rest frames of
$\stau^\pm$. Note that we have already integrated analytically over
the phase space of the invisible decay products (neutrinos and
neutralinos). This fixes the integration limits for $E^*_{\pi^\pm}$
as
\be \label{e16}
\left. E^*_{\pi^\pm} \right|_{\rm min,max} = \frac {m_{\ttau^\pm}} {4}
 \left[ \left( 1 + \frac {m_\tau^2 - m^2_{\tilde \chi} } { m^2_{\ttau^\pm} }
\right) \left( 1 + \frac {m^2_\pi} {m^2_\tau} \right)
%\right. \nonumber \\ & \left.
 \pm \left( 1 - \frac {m^2_\pi} {m^2_\tau} \right)
\lambda^{1/2} \left( 1, \frac {m^2_\tau} { m^2_{\ttau^\pm} },
\frac {m^2_{\tilde \chi}} {m^2_{\ttau^\pm}} \right) \right],
\ee
with $m_{\ttau^-} = m_{\ttau_{ik}}$ and $m_{\ttau^+} = m_{\ttau_{jl}}$
in eq.(\ref{e15}).

The structure of eq.(\ref{e15}) is quite similar to that of eq.(\ref{e4}).
However, the functions $D^\pi_{ik}$ now describe the entire $\stau^- 
\rightarrow \tau^- \tchi_1^0 \rightarrow \pi^- \nu_\tau \tchi_1^0$
decay chain, including all spin effects:
\beq \label{e17}
D^\pi_{ik} = \frac {2 m_\tau^2} {m^2_\tau -m^2_\pi} & \left[
\left( m^2_{\ttau_{ik}} - m^2_{\tilde \chi} - m^2_\tau \right)
R_{i1} R^*_{k1} 
- m_\tau m_{\tilde \chi} \left( L_{i1} R^*_{k1} + R_{i1} L^*_{k1} \right)
\right. \nonumber \\ & \left.
+ \frac {m^2_\tau} {m^2_\tau - m^2_\pi} \left( R_{i1} R^*_{k1}
- L_{i1} L^*_{k1} \right) \left( 2 m_{\ttau_{ik}} E^*_{\pi^-}
- m^2_{\ttau_{ik}} + m^2_{\tilde \chi} - m^2_\pi \right) \right],
\eeq
where $L_{i1}, R_{i1}$ are the $\tchi_1^0 \stau_i \tau$ couplings that
already appeared in eq.(\ref{e10}). Note that it is not sufficient to
treat $\stau \rightarrow \tau$ decays as incoherent sum of $\stau
\rightarrow \tau_L$ and $\stau \rightarrow \tau_R$ decays. The
interference between left-- and right--handed $\tau$ leptons gives a
contribution of order $m_\tau$, shown in the second term of
eq.(\ref{e17}); because this term involves the product of left-- and
right--handed $\tau$ couplings it allows to produce a CP--odd phase
purely from gauge contributions to the couplings $L$ and $R$. In
contrast, the other terms produce a CP--odd phase only if at least one
factor of the $\tau$ Yukawa coupling appears in the relevant
combination of $L$ and $R$ couplings.\footnote{The CP--odd
contribution to eq.(\ref{e15}) is linear in the pion energies
$E^*_{\pi^\pm}$. It therefore involves the product of the
$E^*_\pi$--independent terms from one $D-$function with the term
linear in $E^*_\pi$ from the other $D-$function.} Of course, the
expectation value of any CP--odd observable is again proportional to
the imaginary part of $A_{12}$.

\setcounter{footnote}{0}
Unfortunately we found that the asymmetry in the $\pi^+$ and $\pi^-$
energy distributions is always quite small. The most easily measured
quantity is the total energy asymmetry,
\be \label{e18}
\bar{A}_{E_\pi} = \frac { \langle E_{\pi^+} \rangle - \langle E_{\pi^-}
\rangle } { \langle E_{\pi^+} \rangle + \langle E_{\pi^-} \rangle },
\ee
where the pion energies are now taken in the laboratory frame. In our
scans of parameter space we did not find a scenario where this asymmetry
exceeds 1\%. The ``differential asymmetry''
\be \label{e19}
A_{E_\pi}(E) = \frac { \frac {d \sigma} {d E_{\pi^+}} (E) -
\frac {d \sigma} {d E_{\pi^-}} (E) }
{ \frac {d \sigma} {d E_{\pi^+}} (E) + \frac {d \sigma} {d E_{\pi^-}} (E) }
\ee
can reach the 10\% level in some bins, but such a large asymmetry
always coincides with a very small differential cross section in the
same bin.  This is illustrated in Fig.~3, which shows the differential
asymmetry (solid) and the differential cross section (dashed,
referring to the scale at the right) for one of the most optimistic
scenarios we found.\footnote{In order to reliably predict small
asymmetries without generating hundreds of millions of events we have
symmetrized our Monte Carlo phase space integration against exchange
of $\pi^+$ and $\pi^-$ momenta, including an exchange $\Theta_{\ttau}
\rightarrow - \Theta_{\ttau}$.  This ensures that the numerically
computed asymmetry vanishes in the absence of CP--violation.} In
particular, we have assumed a 100\% right--handed $e^-$ beam here; for
the given choice of parameters this increases the total cross section
by about 30\%, and increases the asymmetry by more than a factor of 2.
We have applied the \stau\ pair acceptance cuts listed in the second
ref.\cite{8}. This reduces the total accepted cross section by about a
factor of 2, and especially depletes the region of small $E_\pi$, but
has no effect on the asymmetry. Note that the differential asymmetry
changes sign near the value of $E_\pi$ where the differential cross
section is maximal. This should allow one to construct an optimized
CP--odd variable, which shows better sensitivity to CP--odd phases
than the total energy asymmetry (\ref{e18}) does, which amounts to
only 0.77\% in the example shown. Nevertheless the significance of the
observed effect cannot exceed that obtained from simply adding the
effective asymmetry observed in each bin in quadrature. This gives a
combined effective asymmetry $\hat{A}_{E_\pi}$ in analogy to the
effective rate asymmetries introduced in the previous section:
\be \label{e20}
\hat{A}^2_{E_\pi} = \int d E A^2_{E_\pi}(E) \left[ \frac {d \sigma}
{d E_{\pi^+}} + \frac {d \sigma} {d E_{\pi^-}} \right].
\ee
Even for the optimistic case depicted in Fig.~3 $\hat{A}_{E_\pi}$ only
amounts to 0.017 fb$^{1/2}$. This compares favorably to the product of
the overall energy asymmetry (\ref{e18}) and the square root of the
total cross section, which only gives 0.0032 fb$^{1/2}$. However, even
if we include events where only one of the two $\tau$ leptons decays
into the single pion mode, which increases the total available cross
section by a factor of $1/B(\tau^- \rightarrow \pi^- \nu_\tau) \simeq
9$ if the efficiencies for the other decay modes are similar to that
for the $\pi^+ \pi^-$ mode, we would still need to accumulate about
400 fb$^{-1}$ of data to see an energy asymmetry at the level of one
standard deviation. In contrast, the same choice of parameters yields
an effective rate asymmetry $\hat{A}( \tchi_1^0 \tchi_1^+ ) = 0.32$
fb$^{1/2}$, which would start to become visible after 20 fb$^{-1}$
have been collected (assuming an efficiency $\epsilon \simeq 50\%$).

\section*{4) Summary and Conclusions}

In this note we have studied CP--violating phenomena that could arise from
scalar $\tau$ oscillations at future \epem\ colliders. In Sec.~2 we found
that rate asymmetries, which can occur if \stau\ has several different
decay modes, can be sizable. However, several conditions have to be
satisfied in order to get asymmetries that might be detectable at a
next--generation collider. The most critical requirement is that the
mass difference between the two \stau\ eigenstates should not exceed 1\%.
In addition, one needs substantial higgsino/gaugino mixing in the
chargino/neutralino sector of the theory and, of course, some significant
CP--odd phases.

Note that the first of these conditions is almost impossible to
fulfill for scalar $b$ or $t$ quarks, since here the off--diagonal
entries in the relevant mass matrices are (much) larger. One might hope
that in case of $\tilde{b}$ squarks at least their larger decay width
could compensate for a larger mass splitting, if $\tilde{b} \rightarrow
\tilde{g} + b$ decays are allowed which proceed through strong interactions.
However, in order to construct a rate asymmetry, at least one of the
$\tilde{b}$ squarks must undergo a weak decay, with correspondingly reduced
branching ratio. Moreover, determining the charges of the final state 
particles, which is crucial for constructing any CP--odd observable, is
not easy for (s)quarks. We therefore do not expect third generation squark
oscillations to be detectable at \epem\ colliders.

In Sec.~3 we studied asymmetries in the pion energy distribution in
events where both \stau\ decay as $\stau \rightarrow \tau \tchi_1^0
\rightarrow \pi \nu_\tau \tchi_1^0$. Unfortunately we found this
asymmetry to be quite small even in the most optimistic case. The
reason is that it vanishes in the limit where the tau mass and Yukawa
coupling are set to zero. This is also true for the rate asymmetries
discussed above; however, in that case substantial cancellations can
and frequently do occur between different contributions to the
relevant \stau -- chargino/neutralino -- $\nu_\tau/\tau$ couplings,
which enhances the relative importance of the $\tau$ Yukawa coupling,
as illustrated in Fig.~1. No such enhancement occurs in the pion
energy asymmetry. The differential energy asymmetry has to change sign
for some value of the pion energy, since the total decay width of the
$\tau$ lepton is independent of its polarization.  The structure of
the decay amplitude implies that this change of sign occurs near the
value of pion energy where the cross section is maximal; as a result,
most events contribute only little to the total asymmetry. At least
for the model with minimal particle content (the MSSM with general
soft breaking terms) studied here, the energy asymmetry therefore
remains too small to be detectable at next--generation \epem\
colliders. This remains true even if we allow for polarized beams,
which can increase this asymmetry by more than a factor of two. On the
other hand, at least in some regions of parameter space rate
asymmetries should be detectable, and might provide us with the first
evidence for CP--violation in the leptonic sector.

\subsection*{Acknowledgements}
The work of SYC was supported in part by the KOSEF--DFG large collaboration
project, Project. No. 96-0702-01-01-2. MD wishes to acknowledge financial
support from the Korean Research Foundation made in the program year of
1997.

%\end{document}

\clearpage
\noindent
\vspace*{2cm}

\setcounter{figure}{0}

\begin{figure}[htb]
\centerline{\epsfig{file=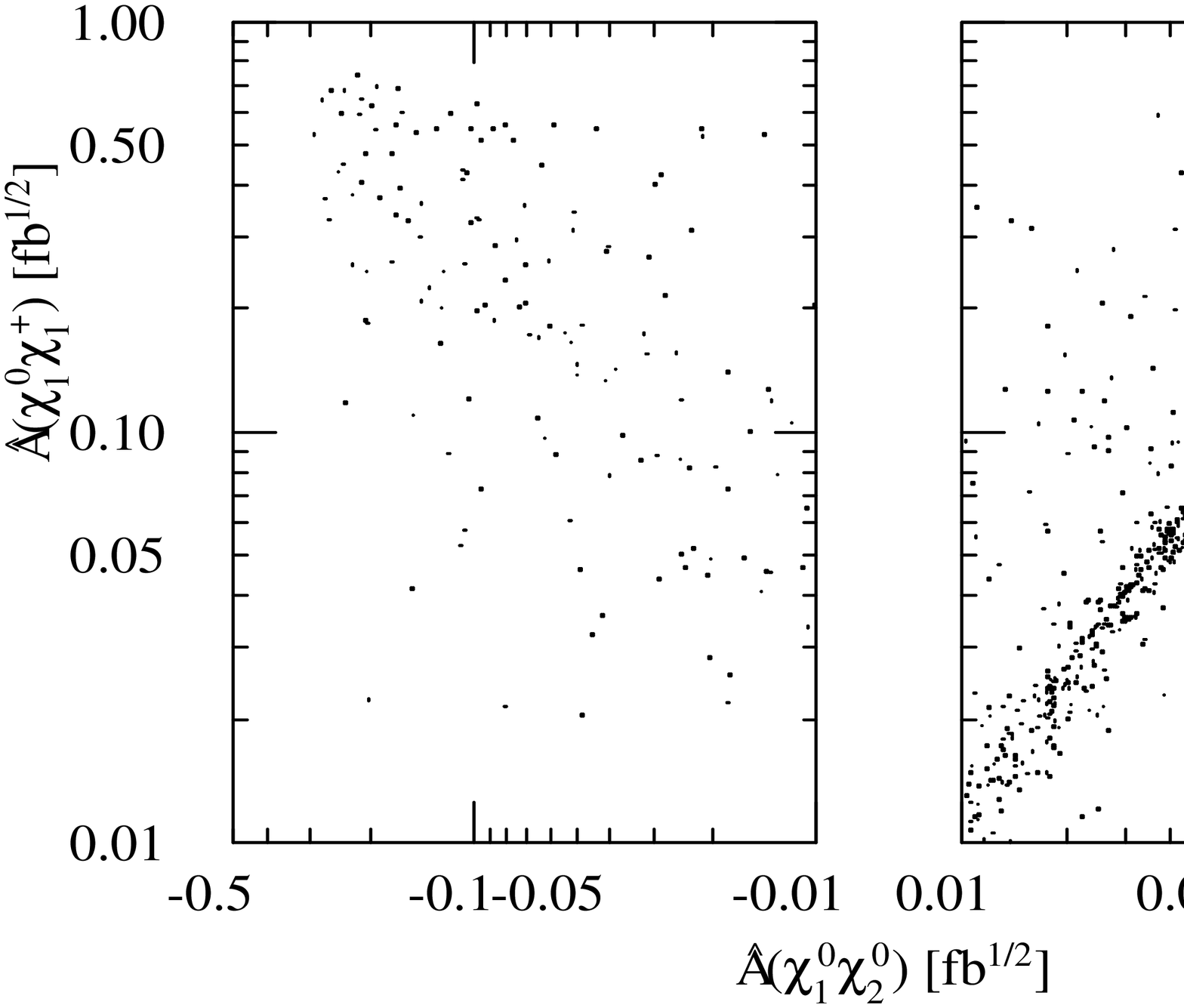,height=9.cm}}

\caption
{Scatter plot of the effective rate asymmetries defined in eq.(13).
We have fixed $m_{\ttau_L} = m_{\ttau_R}=200$ GeV, $|A_\tau| = |\mu
\tanb|$, and $\theta_A = \theta_\mu - 3.2$. The phases of the gaugino
masses have been set to zero. The remaining parameters have been
varied randomly, in the ranges 100 GeV $\leq |M_2|, |\mu| \leq$ 500
GeV, $1 \leq \tanb \leq 5$ and $- \pi \leq \theta_\mu \leq \pi$, where
$\theta_A$ and $\theta_\mu$ are the phases of $A_\tau$ and $\mu$,
respectively. These results are for an \epem\ collider operating at
$\sqrt{s} = 500$ GeV with unpolarized beams.}
\end{figure}

\clearpage
\vspace*{2cm}

\begin{figure}[htb]
\centerline{\epsfig{file=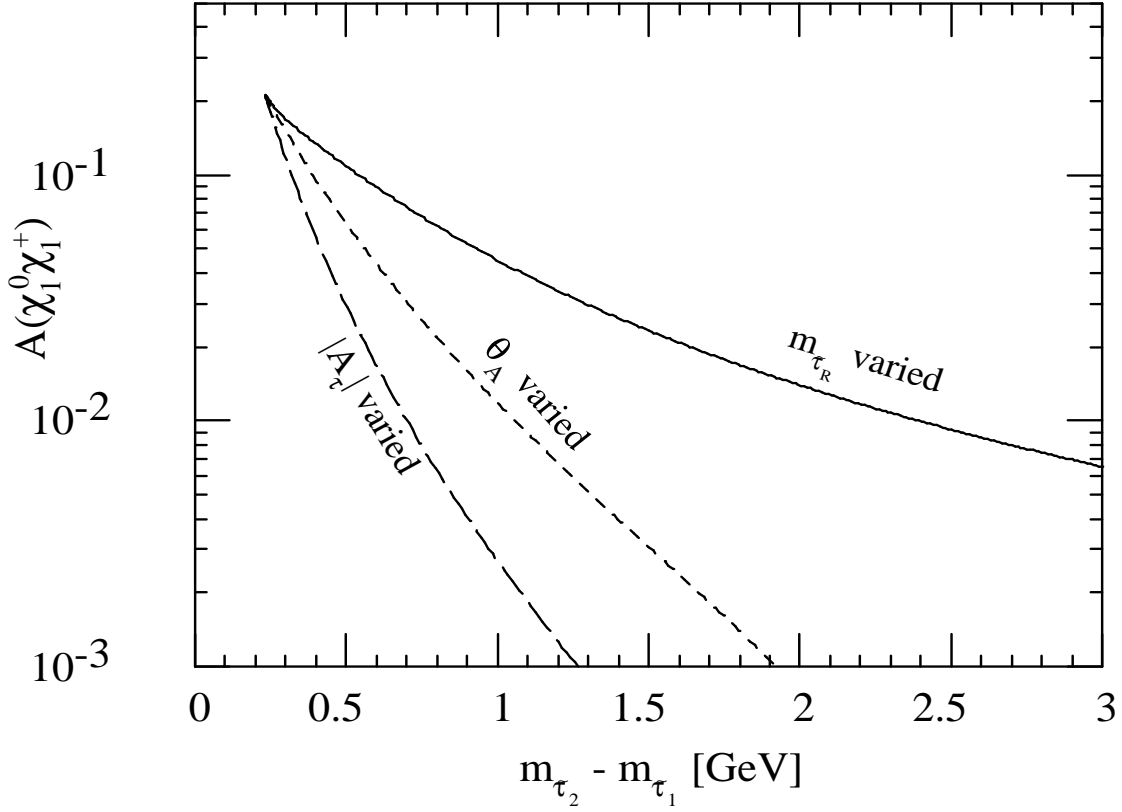,height=9.2cm}}

\caption
{Dependence of the rate asymmetry $A( \tchi_1^0 \tchi_1^+ )$ defined
in eq.(12) on the $\stau_2 - \stau_1$ mass difference. The point
common to all three curves is defined by the parameters $\sqrt{s} =
500$ GeV, $m_{\ttau_L} = m_{\ttau_R} = 200$ GeV, $|A_\tau|=442.6$ GeV,
$|M_2| = 240.2$ GeV, $|\mu|=154.4$ GeV, $\tanb=2.89$, and $\theta_\mu=
\theta_A + 3.2 = 0.19$. Gaugino masses are assumed to be real and to
satisfy a unification condition. The solid, short dashed and long
dashed curves have been obtained by varying $m_{\ttau_R}, \ \theta_A$
and $|A_\tau|$, respectively, as described in the text.}

\end{figure}

\clearpage

\vspace*{2cm}

\begin{figure}[htb]
\centerline{\epsfig{file=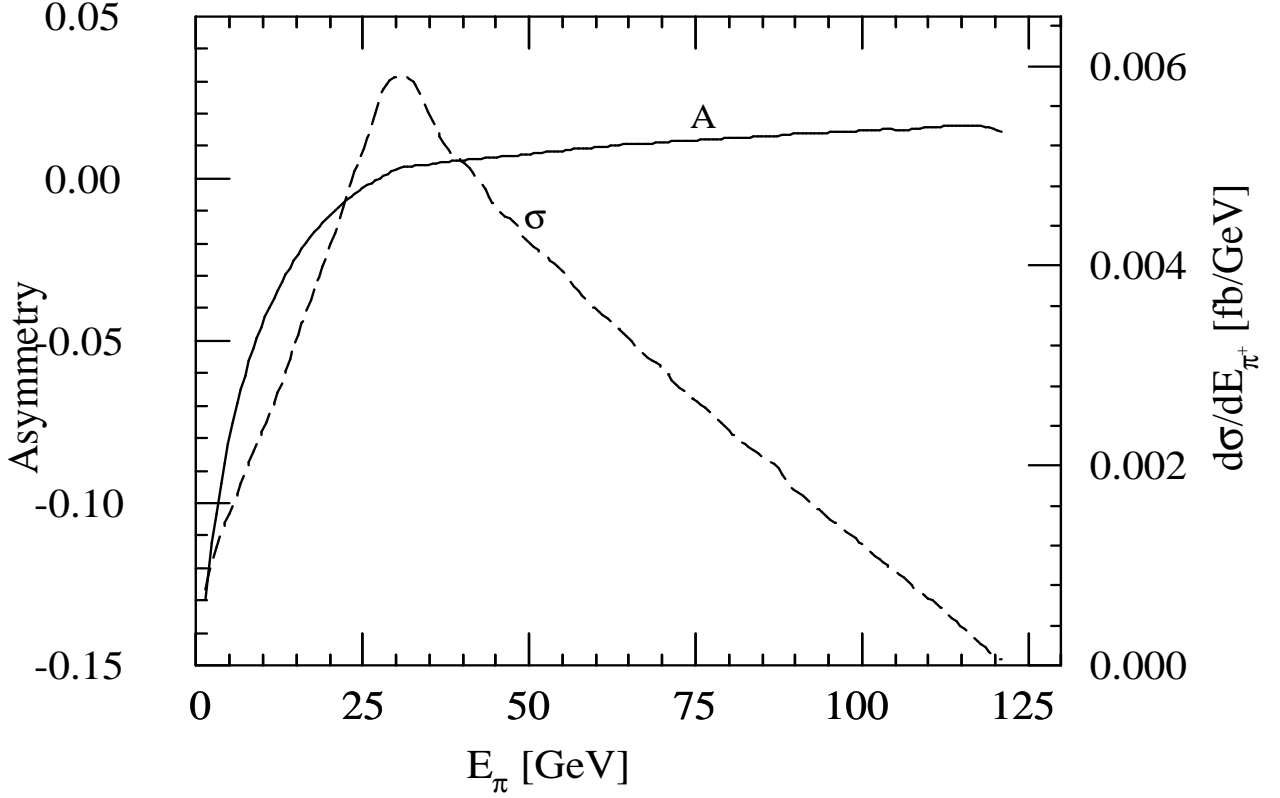,height=9cm}}
\caption
{The differential pion energy asymmetry defined in eq.(19) (solid),
and the differential cross section after acceptance cuts [9]
(dashed); the latter refers to the
scale at the right. Note that both $\tau$ leptons are assumed to decay
into the single pion mode, which gives a combined branching ratio of
only 1.2\%; however, events where only one of the $\tau$ leptons
decays into this mode can also be used for this measurement.  The
values of the relevant parameters are: $|A_\tau| = 454.6$ GeV, $|M_2|
= 270.1$ GeV, $|\mu| = 152.3$ GeV, $\tanb= 2.98$, $\theta_A =
-0.0358$, and $\theta_\mu = 3.1642$; the other parameters are as in
Fig.~2, except that we have used an $e^-$ beam with 100\% right--handed
polarization. }

\end{figure}

\end{document}